\documentclass[preprint]{elsarticle}

\usepackage{array,xspace,multirow,hhline,tikz,colortbl,tabularx,booktabs,fixltx2e,amsmath,amssymb,amsfonts,amsthm}
\usepackage{algorithm}
\usepackage{algorithmic}
\usepackage{verbatim,ifthen}
\usepackage{enumitem}
\usepackage{pifont}
\usepackage{ifthen}
\usepackage{calrsfs,mathrsfs}
\usepackage{bbding,pifont}
\usepackage{pgflibraryshapes}
\usepackage{nicefrac}

\definecolor{light-gray}{gray}{0.9}

	\newtheorem{lemma}{Lemma}%
	\newtheorem{theorem}{Theorem}%
	\newtheorem{example}{Example}

    \newtheorem*{theorem*}{Theorem}

		\newcommand{\ml}[1][]{\ifthenelse{\equal{#1}{}}{\mathit{ML}}{\mathit{ML}(#1)}}
		\newcommand{\sml}[1][]{\ifthenelse{\equal{#1}{}}{\mathit{SML}}{\mathit{SML}(#1)}}
		\newcommand{\sd}[1][]{\ifthenelse{\equal{#1}{}}{\mathit{SD}}{\mathit{SD}(#1)}}
		\newcommand{\rsd}[1][]{\ifthenelse{\equal{#1}{}}{\mathit{RSD}}{\mathit{RSD}(#1)}}
		\newcommand{\st}[1][]{\ifthenelse{\equal{#1}{}}{\mathit{ST}}{\mathit{ST}(#1)}}
		\newcommand{\bd}[1][]{\ifthenelse{\equal{#1}{}}{\mathit{BD}}{\mathit{BD}(#1)}}
		\newcommand{\pc}[1][]{\ifthenelse{\equal{#1}{}}{\mathit{PC}}{\mathit{PC}(#1)}}
		\newcommand{\dl}[1][]{\ifthenelse{\equal{#1}{}}{\mathit{DL}}{\mathit{DL}(#1)}}
		\newcommand{\ul}[1][]{\ifthenelse{\equal{#1}{}}{\mathit{UL}}{\mathit{UL}(#1)}}

		\newcommand{\serdict}[1][]{\ifthenelse{\equal{#1}{}}{\sigma}{\sigma(#1)}}

	\newcommand\eat[1]{}

	\usepackage{enumitem}
	\setenumerate[1]{label=\rm(\it{\roman{*}}\rm),ref=({\it\roman{*}}),leftmargin=*}
	\newlength{\wordlength}

	\newcommand{\set}[1]{\{#1\}}
	\newcommand{\midd}{\mathbin{:}}

	\newcommand{\eqclass}[2][]{\ifthenelse{\equal{#1}{}}{[#2]}{[#2]_{\sim_{#1}}}}

	\newcommand{\pref}{\succsim\xspace}
	\newcommand{\Pref}[1][]{
		\ifthenelse{\equal{#1}{}}{\mathrel R}{\mathop{R_{#1}}}
	}                                          
	\newcommand{\sPref}[1][]{                  
		\ifthenelse{\equal{#1}{}}{\mathrel P}{\mathop{P_{#1}}}
	}                                          
	\newcommand{\Indiff}[1][]{                 
		\ifthenelse{\equal{#1}{}}{\mathrel I}{\mathop{I_{#1}}}
	}
	\newcommand{\prefset}[1][]{\ifthenelse{\equal{#1}{}}{\mathcal{R}}{\mathcal{R}_{#1}}}

\newcommand\Color[1] {\color{#1}}
\definecolor{PurplePlum}{rgb}{0.1,0,0.55} 
\definecolor{Brown}{rgb}{0.5,.25,0}
\definecolor{Green}{rgb}{0,.5,0}
\definecolor{Orange}{rgb}{1,.5,0}
\definecolor{Gray}{rgb}{0.5,0.5,0.5}
\definecolor{Black}{rgb}{0,0,0}

\usepackage{color} 
\usepackage[normalem]{ulem} 

\newcommand\CRx {\Color{Green}}

\newcommand\CRf[2][] 
	{\mbox{\CRx\tiny$\boxtimes$}\marginpar{\CRx\fbox{\FnSym\footnotemark}}
	\footnotetext{\CRx\textsf{#1} #2}}
\newcommand\CRff[2][]
	{\def\File{#1}
	\ifx\File\empty
		\mbox{\CRx\tiny$\boxtimes$}\marginpar{\CRx\fbox{\FnSym\footnotemark}}
		\footnotetext{\CRx\textsf{#1} #2}
	\else
		\mbox{\scriptsize\CRx$\boxtimes$}\marginpar{\CRx\fbox{\FnSym\footnotemark}}
		\footnotetext{\CRx#2\ \EnSym\endnotemark.}
		\endnotetext{\label{#1}\CRx\input{#1}\newpage}
	\fi}
\newcommand\CRe[1] 
	{\mbox{\scriptsize\CRx$\boxtimes$}\marginpar{\CRx\fbox{\EnSym\endnotemark}}\endnotetext{\CRx\ #1}}

\begin{document}
	\title{Incompatibility of Efficiency and Strategyproofness in the Random Assignment Setting with Indifferences}

	 \author{Haris Aziz} \ead{haris.aziz@data61.csiro.au}
      \author{Pang Luo} \ead{pang.luo@student.unsw.edu.au}
		\author{Christine Rizkallah} \ead{christine.rizkallah@data61.csiro.au}

	\address{Data61 and UNSW} 
	


	\begin{abstract}
A fundamental resource allocation setting is the random assignment problem in which agents express preferences over objects that are then randomly allocated to the agents. In 2001, Bogomolnaia and Moulin 
presented the probabilistic serial (PS) mechanism that is an anonymous, neutral, Pareto optimal, and weak strategyproof mechanism when the preferences are considered with respect to stochastic dominance. The result holds when agents have strict preferences over individual objects. It has been an open problem whether there exists a mechanism that satisfies the same properties when agents may have indifference among the objects. We show that for this more general domain, there exists no extension of PS that is ex post efficient and weak strategyproof. The result is surprising because it does not even require additional symmetry or fairness conditions such as anonymity, neutrality, or equal treatment of equals. Our result further demonstrates that the lack of weak SD-strategyproofness of the extended PS mechanism of Katta and Sethuraman (2006) is not a design flaw of extended PS but is due to an inherent incompatibility of efficiency and strategyproofness of PS in the full preference domain. 
\end{abstract}


\maketitle

\section{Introduction}

In the assignment problem, agents express a complete and transitive set of preferences over objects and objects are divided among agents according to these preferences. The problem models one of the most fundamental settings in computer science and economics with numerous applications~\citep{Gard73b,Youn95b,Sven94a,Sven99a,BEL10a,ACMM05a}. 
Depending on the application setting, the objects could be car-park spaces, dormitory rooms, kidneys, school seats, etc.
The assignment problem is also referred to as \emph{house allocation}~\citep{ACMM05a,AbSo99a}.

How do we identify desirable assignment rules for the problem? A natural way is to consider efficiency and strategyproofness with respect to the stochastic dominance (SD) relation. SD is a fundamental way to extend ordinal preferences over individual objects to random allocation because one allocation is SD preferred over another if it yields more utility with respect to all cardinal utility functions consistent with the ordinal preferences.  We consider two requirements --- {ex post efficiency} and weak SD-strategyproofness.  SD-efficiency is Pareto optimality with respect to the SD relation. It is a weak property since it is only violated if the improving agent get more utility with respect to \emph{all} utility functions consistent with the ordinal preferences. Similarly, weak SD-strategyproofness is also a weak property since it is only violated if an agent can misreport their preference and  get more utility with respect to \emph{all} utility functions consistent with the ordinal preferences.

        For the assignment problem, many existing papers assume that the agents have strict preferences over individual objects.
        Although strictness of preferences is natural restriction, it cannot model preferences in which an agent is completely indifferent among some objects because they have the same quality that the agent cares about. 
The most famous mechanism for the problem is \emph{random serial dictatorship (RSD)}: a permutation of the agents is chosen uniformly at random and then agents in the permutation are given the most preferred object that is still not allocated. Although RSD is strategyproof in the strongest sense and also ex post efficient (the outcome can be represented as convex combination of deterministic Pareto optimal outcomes), \citet{BoMo01a} showed that RSD is not SD-efficient even for strict preferences where SD-efficiency is a stronger property than ex post efficiency. 
         They proposed a rival mechanism called \emph{probabilistic serial (PS)} that is anonymous, neutral, SD-efficient, and weak SD-strategyproof.
Under PS, agents `eat' the most favoured available object at an equal rate until all the objects are consumed. When a most preferred object is completely eaten, agents eat their next most preferred object that is not completely eaten. The fraction of object consumed by an agent is the probability of the agent getting that object.

\citet{BoMo01a} left open the problem of generalizing PS for the full domain in which agents may express indifference between objects.
The full domain is a generalization of strict preferences that can also capture other well-studied preferences restrictions such as dichotomous or trichotomous preferences in which agents puts the objects in two or three preference classes.
\citet{KaSe06a} proposed an extension of PS called EPS that is anonymous, neutral, and SD-efficient. However they showed that EPS is not weak SD-strategyproof. Since, the work of \citet{KaSe06a}, it has been open whether there exists an anonymous,  SD-efficient, and weak SD-strategyproof random assignment mechanism. Since, PS satisfies the three properties for strict preferences, it is tempting to think that a rule that satisfies anonymity, SD-efficiency, and weak SD-strategyproofness would be some other interesting extension of PS. A random assignment rule is \emph{an extension of PS} if it returns the same assignment as PS for all preference profiles in which the preferences of each agent are strict. However we prove the following which is the main result of this paper.


    \begin{theorem*}
        For $n\geq 3$, there exists no extension of the probabilistic serial rule that is ex post efficient and weak SD-strategyproof. 
        \end{theorem*}
        
        The impossibility is startling because it does not even require any fairness conditions such as anonymity, neutrality, or even equal treatment of equals.

\paragraph{Related Work}

\citet{BoMo01a} popularize the use of stochastic dominance to define efficiency and strategyproofness in probabilistic settings.
They proved that even for preferences in the strict domain, there exists no random assignment rule that is SD-efficient, SD-strategyproof, and satisfies equal treatment of equals. \citet{KaSe06a} proved that there exists no random assignment rule that is SD-efficient, weak SD strategyproof, and satisfies SD envy-freeness. \citet{Aziz14d} proved that when there are more objects than agents, there exists no anonymous, neutral, SD-efficiency, and weak SD-strategyproof random assignment rules. It still remains to be proven whether there exists a random assignment rule that is anonymous, weak SD-strategyproof, and SD-efficient for equal numbers of agents and objects. 

%
%
%
%
%
%

\citet{BBG15a} have recently proved that for the randomized voting setting, there exists no anonymous, neutral, SD-efficient, and weak SD-strategyproof rule settling a conjecture of \citet{ABBH12a}. However their result does not imply that there exists no such rule for a random assignment problem since the random assignment problem is more restricted and structured than voting.

\section{Preliminaries}

The model we consider is the \emph{random assignment problem} which is a triple $(N,O,\pref)$ where $N$ is the set of $n$ agents $\{1,\ldots, n\}$, $O=\{o_1,\ldots, o_n\}$ is the set of objects, and $\pref=(\pref_1,\ldots,\pref_n)$ is a preference profile specified by a tuple of complete and transitive preference relations $\pref_i$ of agent $i$ over objects in $O$. A good reference for this setting is \citep{BoMo01a}. 
We will denote by $\mathcal{R}(O)$ the set of all complete and transitive relations over the set of objects $O$.

A random assignment $p$ is an $n\times n$ matrix $[p(i)(o_j)]_{1\leq i\leq n, 1\leq j\leq n}$ such that for all $i\in N$, and $o_j\in O$, $ p(i)(o_j) \in [0,1]$; $\sum_{i\in N}p(i)(o_j)= 1$ for all $j\in \{1,\ldots, n\}$; and $\sum_{o_j\in O}p(i)(o_j)= 1$ for all $i\in N$. 
The value $p(i)(o_j)$ represents the probability of object $o_j$ being allocated to  agent $i$. Each row $p(i)=(p(i)(o_1),\ldots, p(i)(o_n))$ represents the allocation of agent $i$. 
The set of columns correspond to probability vectors of the objects $o_1,\ldots, o_n$.
A random assignment is \emph{discrete} if $p(i)(o)\in \{0,1\}$ for all $i\in N$ and $o\in O$.

A \emph{random assignment rule} specifies for each preferences profile a random assignment.
Two minimal fairness conditions for rules are \emph{anonymity} and \emph{neutrality}. A rule is anonymous if its outcome depends only on the  preference profile and does not depend on the identity of the agents. A rule is neutral if its outcome depends only on the preference profile and does not depend on the identity of the objects. A rule satisfies equal treatment of equals if agents with identical preferences get identical allocations. Note that anonymity implies equal treatment of equals.

In order to reason about preferences over random allocations, we extend preferences over objects to preferences over random allocations. One standard extension is \emph{SD (stochastic dominance)}. 
Given two random assignments $p$ and $q$, it holds that $p(i) \succsim_i^{\sd} q(i)$ i.e.,  a player $i$ \emph{$\sd$~prefers} allocation $p(i)$ to allocation $q(i)$ if for all $o\in O$: 
	\[	\sum_{o_j\in \set{o_k\midd o_k\succsim_i o}}p(i)(o_j) \ge \sum_{o_j\in \set{o_k\midd o_k\succsim_i o}}q{(i)(o_j)}.\] Note that  SD is incomplete with respect to allocations, hence, it can be the case that two allocations $p(i)$ and $q(i)$ are \emph{incomparable}: $p(i) \not\succsim_i^{\sd} q(i)$ and  $q(i) \not\succsim_i^{\sd} p(i)$.

	An assignment $p$ is \emph{$\sd$-efficient} if there exists no assignment $q$ such that $q(i) \succsim_i^{\sd} p(i)$ for all $i\in N$ and $q(i) \succ_i^{\sd} p(i)$ for some $i\in N$. A discrete assignment is Pareto optimal if and only if it is $\sd$-efficient. An assignment is \emph{ex post efficient} if it can be represented as a probability distribution over the set of Pareto optimal discrete assignments.

	A random assignment function $f$ is \emph{$\sd$-strategyproof} if 
	$f(\pref)(i)\succsim_i^{\sd} f(\pref_i',\pref_{-i})(i)\text{ for all $\pref_i'\in \mathcal{R}(O)$ and $\pref_{-i}\in {\mathcal{R}(O)}^{n-1}$}.$
	A random assignment function $f$ is \emph{weak $\sd$-strategyproof} if 
	$\neg[f(\pref_i',\pref_{-i})(i) \succ_i^{\sd}  f(\pref)(i)]  \text{ for all $\pref_i'\in \mathcal{R}(O)$ and $\pref_{-i}\in {\mathcal{R}(O)}^{n-1}$}.$
It is easy to see that  $\sd$-strategyproofness implies weak $\sd$-strategyproofness~\citep{BoMo01a}.

The following is an example of a random assignment problem in the full  preference domain, along with an example of a random assignment rule to solve it: 
		\begin{example}[Illustration of the random assignment problem]
		Consider the following random assignment problem where $N=\{1,2,3\}$, $O=\{o_1,o_2,o_3\}$.
		
		The preferences are:
		
		\begin{align*}
			\pref_1:&\quad o_1,\{o_3,o_2\}\\
			\pref_2:&\quad o_2,o_1,o_3\\
			\pref_3:&\quad o_2,o_1,o_3
			\end{align*}

					\[
					x=\begin{pmatrix}
					0.99&0&0.01\\
					0&0.99&0.01\\
					 0.01&0.01&0.98\\
					   		 \end{pmatrix}.
					\]
                    
    Agent 1 most prefers $o_1$ and then is indifferent between $o_2$ and $o_3$. Assignment $x$ specifies how much fraction of an object an agent gets. For example, agent $1$ gets $0.99$ of $o_1$.
			\end{example}

RSD is a random assignment rule in which a permutation of agents is chosen uniformly at random and agents in the permutation successively take their most preferred available object~\citep{AbSo98a,BoMo01a,ABB13b}.


%
%
%

\section{Result}

For strict preferences, PS is the only known assignment rule that is anonymous, SD-efficient, and weak SD-strategyproof. EPS is the only known generalization of PS~\citep{KaSe06a} to the case of indifferences and it is anonymous and SD-efficient but not weak SD-strategyproof. In view of these facts, it is tempting to conjecture that a rule that satisfies anonymity, SD-efficiency, and weak SD-strategyproofness would be an extension of PS.  A random assignment rule is an \emph{extension of PS} if it returns the same assignment as PS for all preference profiles in which all preferences of all agents are strict. However we prove the following theorem which implies that there is no such extension:  For $n\geq 3$, there exists no extension of the probabilistic serial rule that is ex post efficient and weak SD-strategyproof.

    Before we proceed, we introduce a known graph theoretic characterization of SD-efficiency~ \citep{KaSe06a} that we later on use to present a new result about the relation between SD-efficiency and ex post efficiency. 
 A directed graph $G$ is a pair of vertices $V$ and directed edges $E$ of type $V\times V$. A \emph{path} between two vertices in the graph is a sequence of vertices in $V$ such that each two consecutive vertices $u,v$ are connected by a directed edge $(u,v)$ in $E$. A \emph{cycle} in $G$ is a path such that the first and last vertex are the same. 
  Intuitively given an assignment $p$, we are interested in graphs  where $V \subseteq N \cup O$ and $E \subseteq (O\times N)\cup (N \times O)$, and in ``trading cycles" where objects point to agents they belong to, according to $p$, and agents point to objects they prefer at least as much as the object in the previous edge in the cycle. At least one edge from agent to object has to denote a strict preference. 
An assignment $p$ admits a \emph{trading cycle} $o_0,i_0,o_1,i_1,\ldots, o_{k-1},i_{k-1},o_0$ in which $p(i_j)(o_j)>0$ for all $j\in \{0,\ldots, k-1\}$, $o_{j+1 \mod k} \succsim_j o_{j\mod k}$ for all $j\in \{0,\ldots, k-1\}$, and  $o_{j+1 \mod k} \succ_j o_{j\mod k}$ for some $j\in \{0,\ldots, k-1\}$. 
We say the cycle has \emph{size} $k$ because there are $k$ occurrences of agent/object pairs.   \citet{KaSe06a} proved that an assignment is SD-efficient if and only if it does not admit a trading cycle. Next, we identify an interesting insight about trading cycles that is useful for our result and may also be of general interest.

    \begin{lemma}\label{lemma:cyclesize}
If there exists a trading cycle, then there exists a trading cycle of size $k\leq n$ in which there are $k$ unique agents and $k$ unique objects. 
        \end{lemma}
        \begin{proof}
            We show that if there exists a trading cycle in which some object is repeated then there exists a trading cycle of size at most $n$ in which no object is repeated.
            Consider a trading cycle using which the allocation of agent $i$ strictly improves. Let the occurrence of $i$ in which $i$ points to a strictly preferred object (than the object that points to $i$)  be labeled as $i^*$.
    Now consider some object $o$ that is repeated multiple times in the cycle. As we follow the cycle starting from $i^*$,  consider the first occurrences of $o$. There is some agent $j$ who is pointing to $o$. Then make $j$ point to the last occurrences of $o$ before $i^*$. This means that there is a shorter cycle in which $i^*$ strictly improved but there is only one occurrence of object $o$. By repeating such operations, we can obtain a trading cycle in which each object, that occurred in the initial trading cycle,  occurs exactly once. 
   
   \bigskip 
  
Now we show that if there is a trading cycle in which some agent is repeated then there exists a trading cycle in which no agent is repeated. We distinguish between two types of agents. We first consider an agent who in at least one occurrence, points to a strictly preferred object than the object pointing to it. In the second case, we will handle agents who do not satisfy such a condition. 
\begin{enumerate}
    \item
    We first consider an agent $i$ that in at least one occurrence points to a strictly preferred object than the object pointing to it. We denote such an occurrence of $i$ as $i^*$, the object that $i^*$ points to as $o^*$, and the object that points to $i^*$ as $o^{**}$. We know $o^*\succ_i o^{**}$. 
    We distinguish two cases depending on whether $o^*$ is a most preferred object of $i$ in the trading cycle:
    \begin{itemize}
   \item In case $o^*$ is a most preferred object of $i$ in the trading cycle, we check if $o^*$ is the only most preferred object of $i$. We simply let the last occurrence of $i$ before $i^*$ point to $o^*$ and the number of occurrences of $i$ in the trading cycle will be reduced by one. If $o^*$ is not the only most preferred object of $i$, we denote another most preferred object of $i$ as $o^\prime$. We let  $i^*$ point to $o^\prime$ and the number of occurrences of $i$ in the trading cycle will be reduced by at least one.
 \item   For the second case, there must be another occurrence of $i$, denoted as $i^\prime$, other than $i^*$, which points to the most preferred object of $i$. We simply let $i^*$ point to the object that $i^\prime$ points to. In this way, the number of occurrences of $i$ in the trading cycle is reduced by at least one.
    \end{itemize}
    %



\item Now consider an agent $j$ that occurs multiple times and never points to a strictly  preferred object.
By definition, the trading cycle contains at least one occurrence $i^*$ of some agent $i$ that points to a strictly  preferred object. We trace the cycle starting from $i^*$. Let the first occurrence of $j$ be $j_1$ and the last one before $i^*$ be $j_{\ell}$. Now if $j_{1}$ does not own a strictly preferred object than $j_{\ell}$, then $j_1$ simply points to the object pointed by $j_{\ell}$ and we obtain a cycle with only one occurrence of $j$. The smaller cycle is still a trading cycle because it contains $i^*$ that points to a strictly  preferred object than the one it owns. 
If $j_1$ owns a strictly  preferred object than $j_{\ell}$, then we make $j_{\ell}$ point directly to the object that $j_{1}$ points to. We obtain a cycle in which $i^*$ is not present and there is one less occurrence of $j$, namely $j_1$ is not present, and the improving agent is $j$ in particular the occurrence $j_{\ell}$ of $j$. 

\end{enumerate}
By repeating this operation to eliminate repeated occurrences of agents we obtain a trading cycle with no multiple occurrences of agents.

            \end{proof}
    
    \begin{lemma}\label{lemma:expost-sd}
        For $n=3$ and allowing for indifferences, the set of ex post efficient assignments coincides with the set of SD-efficient assignments.
        \end{lemma}
    \begin{proof}
It is well-known that SD-efficiency implies ex post efficiency~\citep{BoMo01a,KaSe06a}. We now show the converse by contradiction for $n=3$. Assume that there exists a random  assignment $p$ that is ex post efficient but not SD-efficient. 
If $p$ is not SD-efficient, then by Lemma~\ref{lemma:cyclesize} it admits a trading cycle of size either two or three in which each agent and object, that occurred in the initial trading cycle, occurs once.

We first assume that there exist agents $1$, $2$, and objects $a$, $b$ such that $p$ admits a trading cycle $b,1,a,2$ of size two in which, without loss of generality, the allocation of $1$ strictly improves by getting $a$ in exchange for $b$.

\[a\succ_1 b\]
\[b\succsim_2 a\]

We now need to show that there exists some Pareto optimal discrete assignment in which $1$ gets $b$ and there exists another in which $2$ gets $a$.
Let us assume that there exists some Pareto optimal discrete assignment in which $1$ gets $b$. This  implies that $2$ gets $c$ and $3$ gets $a$. In order for the discrete assignment to be Pareto optimal, it must be the case that $a\succ_3 b$ or else agents $1$ and $3$ can exchange objects $a$ and $b$.
\[a\succ_3 b\]

Now assume that there exists some Pareto optimal discrete assignment in which $2$ gets $a$. Then $1$ must get $c$ and $3$ gets $b$. But then $3$ can get $a$ from $2$ and give $b$ in return to affect a Pareto improvement. Hence there does not exist any Pareto optimal discrete assignment in which $2$ gets $a$. Thus, there is no trading cycle of size two.
\bigskip 

We now consider the second case in which there exists agents $1$, $2$, $3$, and objects $a$, $b$, $c$ such that $p$ admits a trading cycle $1,a,2,b,3,c,1$ of size three in which, without loss of generality,  the allocation of agent $1$ strictly improves by getting object $a$. The definition of trading cycle implies the following relations.

\[a\succ_1 c\]
\[b\succsim_2 a\]
\[c\succsim_3 b\]

The trading cycle also implies that there exists some Pareto optimal discrete assignment in which $1$ gets $c$, there exists some Pareto optimal discrete assignment in which $2$ gets $a$  and there exists some Pareto optimal discrete assignment in which $3$ gets $b$.
If there exists a Pareto optimal discrete assignment in which $1$ gets $c$, then $2$ gets $b$ and $3$ gets $a$. In order for this discrete assignment to be Pareto optimal it must be the case that 
\[a\succsim_3 c\]
or else $1$ and $3$ can exchange $a$ and $c$ to get a Pareto improvement. 

Now let us consider a discrete Pareto optimal assignment in which $3$ gets $b$. Then it must be that $1$ gets $a$ and $2$ gets $c$. In order for the assignment to be Pareto optimal it must be that 
\[c\succsim_2 b\]
or else $2$ can gets some $b$ from $3$ in exchange of $c$.

Now let us consider a discrete Pareto optimal assignment in which $2$ gets $a$. 
Then it must be that 1 gets $b$ and $3$ gets $c$. But then $3$ can give $c$ to $2$ and get $a$ in return to affect a Pareto improvement. Hence, it cannot be the case that there exists a discrete Pareto optimal discrete assignment such that $1$ gets $c$, $2$ gets $a$, and $3$ gets $b$.
\end{proof}
    
    We have proved that for $n=3$, SD-efficiency and ex post efficiency are equivalent. We use this lemma in the proof of our main result. The lemma also generalizes Lemma 2 (ii) by \citet{BoMo01a} which is limited to strict preference profiles. \citet{BoMo01a} prove the statement by enumerating different classes of preference profiles whereas our proof argument is different as in our case there many more cases to handle. 
    
    \begin{theorem}
        For $n\geq 3$, there exists no extension of PS that is ex post efficient and weak SD-strategyproof. 
        \end{theorem}
        \begin{proof}
            Let us consider a random assignment rule $f$ that is 
            an extension of PS that is ex post efficient, and weak SD-strategyproof for $n=3$. 
            Since ex post efficiency is equivalent to SD-efficiency for $n=3$ (Lemma~\ref{lemma:expost-sd}), we can assume that $f$ is SD-efficient. We will show that no such rule exists by deriving a contradiction. 
            
          We focus on the following three preference profiles $\pref,\pref'$ and $\pref''$. The first two involve only strict preferences.
            
    		\begin{align*}
    			\pref_1:&\quad a,b,c\\
    			\pref_2:&\quad a,c,b\\
    			\pref_3:&\quad a,b,c
    			\end{align*}

    		\begin{align*}
    			\pref_1':&\quad a,b,c\\
    			\pref_2':&\quad a,c,b\\
    			\pref_3':&\quad b,c,a
    			\end{align*}
                
        		\begin{align*}
        			\pref_1'':&\quad a,b,c\\
        			\pref_2'':&\quad a,c,b\\
        			\pref_3'':&\quad \{a,b\},c
        			\end{align*}
                    
                    It can be ascertained that PS outcomes of profiles $\pref$ and $\pref'$ are as follows:
                    
					\[
					PS(\pref)=\begin{pmatrix}
					\nicefrac[]{1}{3}&\nicefrac[]{1}{2}&\nicefrac[]{1}{6}\\
					\nicefrac[]{1}{3}&0&\nicefrac[]{2}{3}\\
					 \nicefrac[]{1}{3}&\nicefrac[]{1}{2}&\nicefrac[]{1}{6}\\
					   		 \end{pmatrix}.
					\]

					\[
					PS(\pref')=\begin{pmatrix}
					\nicefrac[]{1}{2}&\nicefrac[]{1}{4}&\nicefrac[]{1}{4}\\
					\nicefrac[]{1}{2}&0&\nicefrac[]{1}{2}\\
					 0&\nicefrac[]{3}{4}&\nicefrac[]{1}{4}\\
					   		 \end{pmatrix}.
					\]

                    Since $f$ is an extension of PS, it follows that $f(\pref)=PS(\pref)$ and $f(\pref')=PS(\pref')$.
                    Let us assume that the outcome of $f(\pref'')$ is as follows.
                    
					\[
					f(\pref'')=\begin{pmatrix}
					C_{1a}&C_{1b}&C_{1c}\\
					C_{2a}&C_{2b}&C_{2c}\\
					C_{3a}&C_{3b}&C_{3c}\\
					   		 \end{pmatrix}.
					\]
                    
                    We first claim that SD-efficiency of $f$ requires that $C_{3a}=0$. Assume for contradiction that $C_{3a}>0$. SD-efficiency requires that $C_{1b}=0$ and also $C_{2b}=0$. But if  $C_{1b}=C_{2b}=0$, then it implies that $C_{3b}=1$. But this is a contradiction because $C_{3a}>0$.

Since we have established that $C_{3a}=0$, it follows that:
\begin{equation}
     \label{eq=1}
    C_{3b}+C_{3c}=1.
    \end{equation}

                    Weak SD-strategyproofness requires that for all agents each pair of the three profiles is SD-equivalent or incomparable. Otherwise, the SD preferred profile could be considered the misreport that is used to cause an improvement. Both SD-equivalence and incomparability imply the following two conditions for agent 3:
                    \begin{enumerate}
                        \item $f(\pref)(3)\not{\succ_3''}^{SD} f(\pref'')(3)$:
                      \begin{enumerate}
                            \item $C_{3a}+C_{3b}\geq 1/2+1/3=5/6$
                       \end{enumerate}
                       Since $C_{3a}=0$, it follows that  
                       \begin{equation}
                           \label{eq:5/6}
                         C_{3b}\geq 5/6.
                           \end{equation}
                       

                   \item $f(\pref'')(3)\not{\succ_3'}^{SD} f(\pref')(3)$
                          \begin{enumerate}
                       \item $3/4=C_{3b}$ and $1/4=C_{3c}$
                       \item or $3/4>C_{3b}$
                       \item or $1>C_{3b}+C_{3c}$
                \end{enumerate}
                   The first condition (a) cannot hold because
                   of \eqref{eq:5/6}.
                   The second condition (b) also cannot hold because
                   of \eqref{eq:5/6}.
                   The third condition cannot hold because of \eqref{eq=1}. This leads to a contradiction.
                    \end{enumerate}
                    
                    Hence for $n=3$, if $f$ is an extension of PS, and is SD-efficient, then $f$ is not weak SD-strategyproof. 
For $n>3$, we can adapt the same argument by adding agents and their corresponding objects. Each agent exclusively  prefers their corresponding object the most. The original three agents are not interested in these additional objects and have them lower down in their preference list.                    
                    \end{proof}
    

In this paper, we proved that there exists no extension of the PS rule that is ex post efficient and weak SD-strategyproof. 
Previously, \citet{KaSe06a} presented EPS which is a particular extension of PS and showed that it is not weak SD-strategyproof. Our result further demonstrates that the lack of weak SD-strategyproofness is not a design flaw of EPS but is due to an inherent incompatibility of  efficiency and strategyproofness of PS in the full preference domain. 
Moreover, our result is not restricted to rules that employ eating after tie-breaking in case there are indifferences but to \emph{any} rule that only coincides with PS over the strict preference profiles. It still remains to be settled whether there exists an anonymous, SD-efficient, and weak SD-strategyproof random assignment rule. 
	
		\section{Acknowledgments}
	Data61 is funded by the Australian Government through the Department of Communications and the Australian Research Council through the ICT Centre of Excellence Program. The authors thank Nicholas Mattei for sharing his code for the PS mechanism for running some experiments.


\end{document}